\documentclass[prl, twocolumn]{revtex4}

\usepackage{amsmath,amssymb}  
\usepackage{psfrag}  
\usepackage[dvips]{epsfig}  
\usepackage{epsfig}  
\usepackage{color}  

\input psfig.sty

\usepackage{theorem}  

\newtheorem{proposition}{Proposition}  

\theorembodyfont{\rmfamily}  
  
\theoremstyle{break}  
  
\def\QED{~\rule[-1pt]{5pt}{5pt}\par\medskip}  
  
\newenvironment{proof}{{\bf Proof: \ }}{ \hfill \QED}

\def\R{\mathbb{R}}

\def\su{\text{su}}  
\def\dSWAP{\sqrt{\text{SWAP}}} 
\def\SU{\text{SU}}  
 
\begin{document}   
\title{Exact two-qubit universal quantum circuit}

\author{Jun Zhang$^1$, Jiri Vala$^{2, 3}$,  Shankar Sastry$^1$ 
and K. Birgitta Whaley$^{2, 3}$} 
  
\affiliation{$^1$Department of Electrical Engineering  
and Computer Sciences, University of California, Berkeley, CA 94720\\  
$^2$Department of Chemistry and Pitzer Center for Theoretical Chemistry, 
University of California, Berkeley, CA 94720\\  
$^3$Mathematical Sciences Research Institute, 
1000 Centennial Drive, Berkeley, CA 94720} 
 
\date{\today}  
  
\begin{abstract}  
We provide an analytic way to implement any  
arbitrary two-qubit unitary operation, given an 
entangling two-qubit gate together with local gates.  
This is shown to provide explicit  
construction of a universal quantum circuit that 
exactly simulates arbitrary two-qubit operations in $\SU(4)$.   
Each block in this circuit is given in a closed form solution.  
We also provide a uniform upper bound of the applications of
the given entangling gates, and find that exactly 
half of all the Controlled-Unitary gates satisfy the same upper bound
as the CNOT gate.
These results allow for  the efficient implementation of operations in $\SU(4)$ 
required for both quantum computation and quantum simulation.
\end{abstract}   
  
\maketitle  
Construction of explicit quantum circuits that are universal, {\em 
  i.e.}, can implement any arbitrary unitary operation, plays   
a central role in physical implementations   
of quantum computation and quantum   
information processing~\cite{Deutsch:85,Nielsen:00}.  
Despite considerable efforts, there are still very few examples of  
exact universal quantum circuits. 
Most universality results are not constructive, following instead  
the approximative paradigm outlined by Lloyd \cite{Lloyd:95} and   
Deutsch {\it et al.}  \cite{Deutsch:95}, who showed that almost any quantum  
gate for two or more qubits can approximate any desired unitary  
transformation to arbitrary accuracy.  Specific results of Barenco  
{\it et al.} \cite{Barenco:95} showing that a  
combination of quantum Controlled-NOT (CNOT) and single-qubit gates is 
universal in the sense that any unitary operation on arbitrarily many  
qubits can be exactly expressed as a composition of these gates have  
led to the commonly adopted paradigm (``standard model'') of CNOT  
and single-qubit rotations.   
The Brylinski's~\cite{Brylinski:01} showed more generally that  
a two-qubit gate  can provide universality with local gates if  
and only if it is entangling.  
However, this proof is not constructive and does not  
provide exact gate sequences for general operations,  
whereas in practical applications 
it is essential to find a {\em constructive} way to realize the  
two-qubit gates from a given entangling gate and local gates.
Bremner {\it et al.}~\cite{Bremner:02} have recently developed a 
constructive approach to implement the CNOT gate that 
relies on numerical procedures.   

We have previously constructed a quantum circuit that 
contains at most three nonlocal gates generated by a given pure two-body 
Hamiltonian for finite time durations, supplemented with at most four 
local gates~\cite{Zhang:02}.  
Using a geometric theory, we proved that such a quantum circuit can 
simulate any arbitrary two-qubit operation exactly and is therefore universal. 
However, in many physical applications, 
one may have little  flexibility in choice of 
Hamiltonian or time duration.
  In this situation, 
what we can access is often a prescribed entangling gate $U_g$ which is
generated by a Hamiltonian over a fixed time duration.  
We present here an exact analytical approach to construct 
exact universal quantum circuits from such an arbitrary 
given entangling gate together with local gates.  
Our approach is based on the recognition in~\cite{Brylinski:01} that,  
given the group of two-qubit gates with subgroup $H$ of local gates,  
$H'=U_gHU_g^{-1}$ is also a subgroup, 
where $U_g$ is a given entangling gate. It can then be shown that the  
Lie algebras of $H$ and $H'$  
generate $\su(4)$, the Lie algebra of the special unitary group
$\SU(4)$. We develop an analytic realization of $H'$ and use this to
construct an exact quantum circuit for any arbitrary two-qubit
operation in $\SU(4)$.  Each step in the construction of the quantum  
circuit is given in a closed form solution.
 
One of the main features of this work is that we provide a uniform upper 
bound of the applications of the given entangling gate $U_g$, {\it i.e.}, 
regardless of which two-qubit operation is to be implemented, we can 
always construct an exact quantum circuit in which the applications of the given 
entangling gate do not exceed the prescribed number. 
Existence of a uniform upper bound with relatively small number for any given
entangling gate provides an important estimation of
  overhead for experimentalists
considering different physical implementations of two-qubit operations.
The value of this upper bound 
depends solely on the nonlocal part of the given entangling gate. 
Specifically, we find that at most
6 applications of the CNOT gate suffices to implement any 
arbitrary two-qubit operation, and that exactly half of all
Controlled-Unitary gates have the same 
uniform upper bound of 6 applications.  This implies that half of 
the Controlled-Unitary gates can be used 
to implement two-qubit operations just as efficiently as the widely used 
CNOT gate, where efficiency refers to minimizing the 
uniform upper bound to circuit size.  
 Another important feature of this work is that it
suggests a generality beyond the standard model, 
namely, it offers an efficient {\em direct} route to simulate
any arbitrary two-qubit unitary operation with whatever entangling gates arise 
naturally in the physical applications.

\paragraph{Preliminary} 
\label{sec:prelim}  
We first briefly introduce some basic facts about   
Cartan decomposition of $\SU(4)$ and local equivalence of  
two-qubit gates.  
Any two-qubit unitary operation $U\in \SU(4)$ can be decomposed
as~\cite{Khaneja:01,Kraus:01, Zhang:02}  
\begin{equation}  
  \label{eq:1}  
  U=k_1\cdot e^{ c_1\frac{i}{2}\sigma_x^1\sigma_x^2} \cdot  
e^{c_2\frac{i}{2}\sigma_y^1\sigma_y^2} \cdot  
e^{c_3\frac{i}{2}\sigma_z^1\sigma_z^2}\cdot k_2,  
\end{equation} 
where  $\sigma_{\alpha}^1\sigma_{\alpha}^2 =  
 \sigma_{\alpha}\otimes \sigma_{\alpha}$,   
 $\sigma_\alpha$ are the Pauli matrices, and   
$k_1$, $k_2\in {\SU(2)\otimes \SU(2)}$ are local gates.
The tetrahedral representation of nonlocal gates in~\cite{Zhang:02}
defines a unique set of coefficients $c_j$
 satisfying:
\begin{equation}  
  \label{eq:40}  
  \pi-c_2\ge c_1\ge c_2\ge c_3\ge 0.  
\end{equation}  
Also from~\cite{Zhang:02}, we know that local gates  
$U\in \SU(2)\otimes \SU(2)$ correspond to the case when  
$c_1=c_2=c_3=0$, or $c_1=\pi$, $c_2=c_3=0$;  
the SWAP gate to $c_1=c_2=c_3=\frac{\pi}{2}$; and the CNOT gate to  
 $c_1=\frac{\pi}2$, $c_2=c_3=0$.  
  
Two unitary transformations $U$, $U_1\in \SU(4)$ are called \emph{locally  
  equivalent} if they differ only by local operations: $U= k_1 U_1 k_2$,  
  where $k_1$, $k_2\in \SU(2)\otimes \SU(2)$.  
It was shown in~\cite{Zhang:02} that any nonlocal two-qubit  
operation that is not locally equivalent to the SWAP  
gate is entangling. The SWAP gate and its local equivalence  
class are thus the only nonlocal two-qubit operations   
that transform unentangled states to unentangled states, {\em i.e.}, 
that do not introduce any entanglement.  
  
\paragraph{Universal quantum circuit}  
\label{sec:universal}  
We now present an analytic way to   
implement any arbitrary two-qubit gate $U\in \SU(4)$  
by constructing a closed form solution for a universal quantum circuit 
that is  composed of a small number of repetitions of  
a given entangling operation $U_g$ together with local gates.  Local gates 
are assumed to be implementable at ease, as is the case in many of the  
current proposed physical implementations of quantum computation~\cite{Nielsen:00}. 
 
An arbitrary two-qubit operation   
$U\in \SU(4)$ can be written as in Eq. (\ref{eq:1}).   
Letting $k_x=e^{\frac{\pi}4i\sigma_y}\otimes e^{\frac{\pi}4i\sigma_y}$  
and $k_y=e^{\frac{\pi}4i\sigma_x}\otimes e^{\frac{\pi}4i\sigma_x}$,   
we have  
\begin{eqnarray}  
  \label{eq:35}  
  \sigma_x^1\sigma_x^2=k_x^\dag\sigma_z^1\sigma_z^2k_x, \quad 
\sigma_y^1\sigma_y^2=k_y^\dag\sigma_z^1\sigma_z^2 k_y. 
\end{eqnarray}  
Substituting Eq. (\ref{eq:35}) into Eq. (\ref{eq:1}),  
we find that an arbitrary two-qubit gate $U\in \SU(4)$ can be  
written as  
\begin{eqnarray}  
  \label{eq:3}  
U=(k_1k_x^\dag) e^{c_1\frac{i}{2}\sigma_z^1\sigma_z^2} (k_x k_y^\dag)  
e^{c_2\frac{i}{2}\sigma_z^1\sigma_z^2} (k_y)  
e^{c_3\frac{i}{2}\sigma_z^1\sigma_z^2}(k_2),  
\end{eqnarray}  
where $k_1$, $k_2$, $k_x$ and $k_y$ are all local gates.  
Since we have all the local gates at our full disposal, it is evident that  
we only need to implement the nonlocal block 
$e^{c_j\frac{i}2 \sigma_z^1\sigma_z^2}$ from the given entangling gate  
$U_g$ together with local gates, for general values of $c_j$ between 0 and $\pi$. 
We have found the following analytic construction of this general block:  
\begin{description}  
\item[Step 1] Apply $U_g$ at most twice to build a gate   
$e^{\gamma\frac{i}2 \sigma_z^1\sigma_z^2}$ with $\gamma\in (0,
\frac\pi{2}]$ (Proposition~\ref{proposition:ContU}).  The value of 
$\gamma$ obtained depends on the starting $U_g$.
\item[Step 2] Apply $e^{\gamma\frac{i}2 \sigma_z^1\sigma_z^2}$ $n$ times,  
  until $n\gamma \in [\frac\pi{4}, \frac\pi{2}]$.  
\item[Step 3] Apply $e^{n\gamma\frac{i}2 \sigma_z^1\sigma_z^2}$ with  
  $n\gamma\in [\frac\pi{4}, \frac\pi{2}]$   
  twice, to simulate the nonlocal block $e^{c_j\frac{i}2 \sigma_z^1\sigma_z^2}$  
  (Proposition~\ref{proposition:universal}).  
\item[Step 4] Build the quantum circuit according to Eq. (\ref{eq:3}). 
\end{description}  

We now describe this construction in more detail.  
\begin{proposition}  
  \label{proposition:ContU}  
Any arbitrary given entangling gate $U_g$ can simulate a gate  
$e^{\gamma \frac{i}2 \sigma_z^1\sigma_z^2}$, where $\gamma\in (0,  
\frac\pi{2}]$,  by a quantum circuit that  
applies $U_g$ at most twice.  
\end{proposition}  
\begin{proof}  
From Sec.\,a, an arbitrary entangling gate $U_g$ can be uniquely   
written as $U_g=k_l Ak_r$, where  
\begin{eqnarray}  
  \label{eq:2}  
A= e^{ \gamma_1\frac{i}{2}\sigma_x^1\sigma_x^2} \cdot  
e^{\gamma_2\frac{i}{2}\sigma_y^1\sigma_y^2} \cdot  
e^{\gamma_3\frac{i}{2}\sigma_z^1\sigma_z^2},\nonumber 
\end{eqnarray}  
$k_l$, $k_r\in {\SU(2)\otimes \SU(2)}$, and   
$\pi-\gamma_2\ge\gamma_1\ge \gamma_2\ge \gamma_3\ge 0$.   
Since $U_g$ is an entangling gate, {\em i.e.}, it is neither local  
nor locally equivalent to the SWAP gate,   
we do not need to take the following cases into account 
 and may ignore them: (1) $\gamma_1=\gamma_2=\gamma_3=0$; (2) 
$\gamma_1=\pi$, $\gamma_2=\gamma_3=0$; and (3) 
$\gamma_1=\gamma_2=\gamma_3=\frac\pi{2}$.  
The first two cases  
correspond to local gates, and the third one to the SWAP gate.  
For all the remaining possibilities of $\gamma_j$,  
we distinguish the following four cases:  
  
\noindent\emph{Case 1:}
If $\gamma_3=\gamma_2=0$ and $\gamma_1\in (0, \pi)$,    
then $k_x A k_x^\dag=e^{\gamma_1 \frac{i}2 \sigma_z^1\sigma_z^2}$ 
is  in the required form.  
  
\noindent\emph{Case 2:}  
If $\gamma_3=0$ and $\gamma_1=\gamma_2=\frac\pi{2}$, we construct the  
quantum circuit 
\begin{eqnarray}  
  \label{eq:29}  
e^{-\frac\pi{4}i\sigma_y^1}Ae^{\frac\pi{4}i\sigma_y^2}  
e^{\frac\pi{4}i\sigma_z^1 -\frac\pi{4}i\sigma_z^2}  
A e^{-\frac\pi{4}i\sigma_z^1  
 +\frac\pi{4}i\sigma_z^2}e^{\frac\pi{4}i\sigma_y^1}\nonumber 
=e^{\frac\pi{4}i\sigma_z^1\sigma_z^2}.\nonumber 
\end{eqnarray}  
  
\noindent\emph{Case 3:}  
If $\gamma_3=0$, $\gamma_1\in (0, \pi)$, and $\gamma_2\in (0, \frac\pi{2})$,    
we have the quantum circuit as  
$Ae^{\frac{i\pi}2\sigma_x^1}Ae^{-\frac{i\pi}2\sigma_x^1}  
 =e^{2\gamma_1\frac{i}{2}\sigma_z^1\sigma_z^2}$. 
  
\noindent\emph{Case 4:}  
If $\gamma_3 \in (0, \frac\pi{2})$, we use the quantum circuit 
$ Ae^{\frac{i\pi}2\sigma_z^1}Ae^{-\frac{i\pi}2\sigma_z^1} \nonumber 
 =e^{2\gamma_3\frac{i}{2}\sigma_z^1\sigma_z^2}$. 
 
These four cases exhaust all the possibilities of $\gamma_j$.  
We have thus obtained a closed form solution to construction of the gate   
 $e^{\gamma \frac{i}2\sigma_z^1\sigma_z^2}$ from $U_g$ with local  
 gates, where $\gamma\in (0, 2\pi)$ and $\gamma \ne \pi$. 
We can always choose $\gamma\in (0, \pi)$ since  
\begin{eqnarray}  
  \label{eq:38}  
 e^{\gamma\frac{i}{2}\sigma_z^1\sigma_z^2}  
=i\, e^{\frac{\pi i}2\sigma_z^1} 
e^{(\pi+\gamma)\frac{i}{2}\sigma_z^1\sigma_z^2} 
 e^{\frac{\pi i}2\sigma_z^2}. 
\end{eqnarray}  
Furthermore, if $\gamma\in [\frac\pi{2}, \pi)$, we can construct the  
following quantum circuit to bring $\gamma$ into the interval $(0, \frac\pi{2}]$:  
\begin{eqnarray}  
  \label{eq:39}  
-i\, e^{-\frac{\pi i}2\sigma_z^1} 
e^{\frac{\pi i}2\sigma_y^1} 
e^{\gamma\frac{i}{2}\sigma_z^1\sigma_z^2} 
e^{-\frac{\pi i}2\sigma_y^1} 
 e^{-\frac{\pi i}2\sigma_z^2} 
=e^{(\pi-\gamma)\frac{i}{2}\sigma_z^1\sigma_z^2}. 
\end{eqnarray}  
The proof of the Proposition is thereby complete.  
\end{proof}  
  
We have thus shown that given an entangling gate $U_g$ together  
with local gates, we can  
construct a gate $e^{\gamma \frac{i}2 \sigma_z^1\sigma_z^2}$ with 
$\gamma\in (0, \frac\pi{2}]$ in an 
analytic form (Step 1). Now it is evident that if the constructed gate has   
$\gamma\in (0, \frac\pi{4}]$, then it can be applied   
for $n$ times until $n\gamma\in [\frac\pi{4}, \frac\pi{2}]$ (Step 2).   
In the next Proposition, we will use the resulting gate   
 $e^{\gamma \frac{i}2\sigma_z^1\sigma_z^2}$ with   
$\gamma\in [\frac\pi{4}, \frac\pi{2}]$, as a  
basic building block  to simulate any generic nonlocal block 
$e^{c\frac{i}2 \sigma_z^1\sigma_z^2}$ (Step 3). 
From Eqs. (\ref{eq:38}) and (\ref{eq:39}), 
 we only need to consider the case when $c\in (0, \frac{\pi}2]$. 
  
\begin{proposition}  
  \label{proposition:universal}  
Given a gate $e^{\gamma \frac{i}2\sigma_z^1\sigma_z^2}$, where  
$\gamma\in [\frac\pi{4}, \frac\pi{2}]$, together with  
local gates, the following quantum circuit   
can simulate the gate  
 $e^{c \frac{i}2  \sigma_z^1\sigma_z^2}$ for 
 any $c\in (0, \frac{\pi}2]$:  
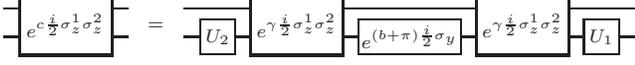
\setlength{\unitlength}{0.15cm}  
\begin{center}  
\scriptsize  
\begin{picture}(56,5)  
\put(0,0){\begin{picture}(16,5)  
\put(0,1.5){\line(1,0){1.5}}  
\put(0,4){\line(1,0){1.5}}  
\put(1.5,0){\thicklines\framebox(8,5)[c]{$e^{c \frac{i}2\sigma_z^1\sigma_z^2}$}}  
\put(9.5,1.5){\line(1,0){1.5}}  
\put(9.5,4){\line(1,0){1.5}}  
\put(11,0){\makebox(5,5)[c]{$=$}}  
\end{picture}}  
\put(16,0){\begin{picture}(40, 5)  
\put(0,1.5){\line(1,0){1.5}}  
\put(1.5,0){\framebox(3,3)[c]{$U_2$}}  
\put(4.5,1.5){\line(1,0){1.5}}  
\put(6,0){\thicklines\framebox(8,5)[c]{$e^{\gamma \frac{i}2\sigma_z^1\sigma_z^2}$}}  
\put(0,4){\line(1,0){6}}  
\put(14,4){\line(1,0){12}}  
\put(14,1.5){\line(1,0){1.5}}  
\put(15.5,0){\framebox(9,3)[c]{$e^{(b+\pi) \frac{i}{2}\sigma_y}$}}  
\put(24.5,1.5){\line(1,0){1.5}}  
\put(26,0){\thicklines\framebox(8,5)[c]{$e^{\gamma\frac{i}2\sigma_z^1\sigma_z^2}$}}  
\put(34,1.5){\line(1,0){1.5}}  
\put(35.5,0){\framebox(3,3)[c]{$U_1$}}  
\put(38.5,1.5){\line(1,0){1.5}}  
\put(34,4){\line(1,0){6}}  
\end{picture}} 
\end{picture}  
\end{center}  
\end{proposition}  
In the above quantum circuit, we have  
\begin{eqnarray}  
\label{eq:9}  
  U_1=\left(\begin{matrix}  
    ip&iq\\ -q&p  
    \end{matrix}\right),\quad  
 U_2=\left(\begin{matrix}  
    ip&-q\\ -iq&-p  
    \end{matrix}\right),\\   
  \label{eq:25}  
 b=\cos^{-1}\big( (\cos{c}-\cos^2\gamma)/\sin^2\gamma\big),  
\end{eqnarray}  
where  
\begin{eqnarray}  
  \label{eq:16}  
  p=\sqrt{\frac12\bigg(1+\frac{\tan{\frac{c}{2}}}{\tan\gamma}\bigg)},\quad  
q=\sqrt{\frac12\bigg(1-\frac{\tan{\frac{c}{2}}}{\tan\gamma}\bigg)}.  
\end{eqnarray}  
\begin{proof}  
We first justify the condition $\gamma\in [\frac\pi{4}, \frac\pi{2}]$. 
From Eq. (\ref{eq:25}), we have $\cos c =\sin^2\gamma \cos b 
+\cos^2\gamma$. Therefore, 
$\cos 2\gamma \le \cos c \le 1$, 
which yields that $0\le c \le 2\gamma$. To cover the full range  
$(0, \frac\pi{2}]$ of $c$, we therefore require that $\gamma\ge \frac\pi{4}$.  
  
We now derive a few formulas required for the proof below.  
It is straightforward to show that $p^2+q^2=1$.  
The identity $\sin\frac{c}{2}=\sin\gamma\sin\frac{b}2$ follows   
from Eq. (\ref{eq:25}) by direct derivations.  
This yields  
\begin{equation}  
  \label{eq:31}  
\tan^2\gamma  
=\frac{\sin^2\frac{c}2}{\sin^2\frac{b}2-\sin^2\frac{c}2}, \nonumber 
\end{equation}  
whence   
\begin{equation}  
  \label{eq:30}  
  pq=\frac12\sqrt{1-\frac{\tan^2{\frac{c}{2}}}{\tan^2\gamma}}  
=\frac{\cos\frac{b}2}{2\cos\frac{c}2}.  
\end{equation}  
  
The Proposition can now be proved. Since $p^2+q^2=1$, it is easy to see that  
$U_1U_1^\dag=U_2U_2^\dag=I$. Hence,  
$U_1$ and $U_2$ are indeed single-qubit gates.  
The quantum circuit can be rewritten as:  
\begin{eqnarray}  
  \label{eq:17}  
&&(I\otimes U_1) e^{\gamma \frac{i}2\sigma_z^1\sigma_z^2}  
(I\otimes e^{(b+\pi)\frac{i}{2}\sigma_y})  
e^{\gamma \frac{i}2\sigma_z^1\sigma_z^2} (I\otimes U_2)\nonumber\\  
&=&\left(\begin{matrix}  
W&\\  
&V  
\end{matrix}\right),  
\end{eqnarray}  
where  
\begin{eqnarray}  
  \label{eq:18}  
W&=& U_1\cdot e^{i\frac{\gamma}2\sigma_z} \cdot  
 e^{(b+\pi)\frac{i}{2}\sigma_y}\cdot e^{i\frac{\gamma}2\sigma_z}\cdot U_2, \\  
\label{eq:32}  
V&=& U_1\cdot e^{-i\frac{\gamma}2\sigma_z} \cdot   
e^{(b+\pi)\frac{i}{2}\sigma_y}\cdot  
e^{-i\frac{\gamma}2\sigma_z}\cdot U_2.  
\end{eqnarray}  
After substituting Eq. (\ref{eq:9}) into Eq. (\ref{eq:18}) and   
applying the identities $\sin\frac{c}{2}=\sin\gamma\sin\frac{b}2$ and  
Eq.~(\ref{eq:30}), we obtain
\begin{eqnarray}  
  \label{eq:19}  
W_{11}&=& p^2\sin\frac{b}2 e^{i\gamma}+2pq\cos\frac{b}2  
-q^2\sin\frac{b}2e^{-i\gamma}=e^{c \frac{i}2},\nonumber\\  
W_{22}&=& p^2\sin\frac{b}2 e^{-i\gamma}+2pq\cos\frac{b}2  
-q^2\sin\frac{b}2e^{i\gamma}=e^{-c \frac{i}2},\nonumber\\  
W_{12}&=&  
W_{21}=2ipq\sin\frac{b}2\cos\gamma-i(p^2-q^2)\cos\frac{b}2=0. 
\nonumber 
\end{eqnarray}  
Hence $W=e^{c \frac{i}2 \sigma_z}$. Similarly,  
we find $V=e^{-c \frac{i}2 \sigma_z}$. Eq. (\ref{eq:17}) now becomes  
\begin{eqnarray*}  
  \label{eq:21}  
&&(I\otimes U_1) e^{\gamma \frac{i}2\sigma_z^1\sigma_z^2}  
(I\otimes e^{(b+\pi)\frac{i}{2}\sigma_y})  
 e^{\gamma \frac{i}2\sigma_z^1\sigma_z^2}(I\otimes U_2)\nonumber\\  
&=&\left(\begin{matrix}  
e^{c \frac{i}2 \sigma_z}&\\  
&e^{-c \frac{i}2 \sigma_z}  
\end{matrix}\right)=e^{c\frac{i}2 \sigma_z^1\sigma_z^2},  
\end{eqnarray*}  
which completes the proof.  
\end{proof}  

Note that in the above Proposition, for the extreme case when  
$\gamma=\frac\pi{2}$, corresponding to starting from a CNOT gate, 
we have $b=c$, and 
\begin{eqnarray}  
  \label{eq:43}  
  U_1=\frac1{\sqrt{2}}\left(\begin{matrix}  
    i&i\\ -1&1  
    \end{matrix}\right),\quad  
 U_2=\frac1{\sqrt{2}}\left(\begin{matrix}  
    i&-1\\ -i&-1  
    \end{matrix}\right). \nonumber 
\end{eqnarray}  

As a physical example, let us consider neutral atoms in an optical lattice 
as a simulator for a solid state many-body spin system. 
The simulation objective may for instance be implementation of
$\dSWAP$.  While this is readily generated 
in spin systems, from the isotropic exchange Hamiltonian
\cite{Burkard:99b}, it is not directly accessible for neutral atoms 
in optical lattices.  A convenient experimentally accessible 
nonlocal transformation in this setting is the Controlled-PHASE gate $C_\phi$, 
where the PHASE gate 
is $\footnotesize\big(\begin{matrix} 1&\\& e^{i\phi}\end{matrix}\big)$. 
From the Cartan decomposition, we have  
\begin{eqnarray} 
  \label{eq:6} 
C_\phi=e^{i\frac{\phi}4}\cdot 
e^{-i\frac{\phi}4\sigma_z}\otimes e^{-i\frac{\phi}4\sigma_z} \cdot 
e^{\frac{\phi}2 \frac{i}2\sigma_z^1\sigma_z^2}. 
\end{eqnarray} 
On the other hand, the $\dSWAP$ gate can be written as:
\begin{eqnarray} 
  \label{eq:8} 
  \dSWAP=e^{-i\frac{\pi}8}\cdot  
e^{\frac{\pi}4 \frac{i}{2}\sigma_x^1\sigma_x^2} \cdot  
e^{\frac{\pi}4 \frac{i}{2}\sigma_y^1\sigma_y^2} \cdot  
e^{\frac{\pi}4 \frac{i}{2}\sigma_z^1\sigma_z^2}.\nonumber 
\end{eqnarray} 
From the quantum circuit in Eq. (\ref{eq:3}), since 
$c_1=c_2=c_3=\frac{\pi}4$, we need only to 
implement the nonlocal gate $e^{\frac{\pi}4 
\frac{i}{2}\sigma_z^1\sigma_z^2}$.  
For $\phi\in [\frac{\pi}2, \pi]$, 
from Proposition~\ref{proposition:universal} and Eq. (\ref{eq:6}), we get
\begin{eqnarray} 
  \label{eq:14} 
  e^{\frac{\pi}4\frac{i}{2}\sigma_z^1\sigma_z^2} 
&=&e^{-i\frac{\phi}2} \cdot 
(e^{i\frac{\phi}4\sigma_z}\otimes U_1 e^{i\frac{\phi}4\sigma_z}) \cdot 
C_\phi \cdot (e^{i\frac{\phi}4\sigma_z} \nonumber\\ 
&&\otimes e^{\frac{i}2(b+\pi)\sigma_y}  
e^{i\frac{\phi}4\sigma_z})\cdot C_\phi \cdot 
(I\otimes U_2), \nonumber 
\end{eqnarray} 
where $ b=\cos^{-1}\big( (\frac{1}{\sqrt{2}}-\cos^2 
\frac{\phi}2)/\sin^2\frac{\phi}2\big)$, $U_1$ and $U_2$ are given as 
in Eq. (\ref{eq:9}), and  
\begin{eqnarray} 
 p=\sqrt{\frac12\bigg(1+\frac{\sqrt{2}-1}{\tan\frac{\phi}2}\bigg)},\quad  
q=\sqrt{\frac12\bigg(1-\frac{\sqrt{2}-1}{\tan\frac{\phi}2}\bigg)}.\nonumber 
\end{eqnarray} 
Thus a spin-spin interaction can be simulated in an optical lattice 
with only 2 repetitions of a Controlled-PHASE gate $C_\phi$ having
 $\phi\in [\frac{\pi}2, \pi]$. 

\paragraph{Uniform upper bound}  
One often desires to simulate arbitrary two-qubit operation by   
applying the given entangling two-qubit  
operation as infrequently as possible.   
From the construction procedure described above, 
we first use the given entangling gate $U_g$ to implement a gate   
 $U_f=e^{\gamma\frac{i}2\sigma_z^1\sigma_z^2}$ with $\gamma\in  
 [\frac\pi{4}, \frac\pi{2}]$ (Proposition~\ref{proposition:ContU}),   
and then apply $U_f$ twice, to implement a 
generic nonlocal gate  
 $e^{c\frac{i}2\sigma_z^1\sigma_z^2}$ 
(Proposition~\ref{proposition:universal}).   
From the decomposition of $\SU(4)$ in Eq.~(\ref{eq:3}),  
any arbitrary two-qubit unitary operation contains at most three 
such nonlocal blocks, resulting in the quantum circuit
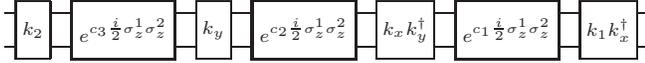
\setlength{\unitlength}{0.15cm}  
\begin{center}  
\begin{picture}(57, 5) 
\scriptsize 
\put(0,0){\begin{picture}(6, 5)  
\put(0,1){\line(1,0){1}}  
\put(0,4){\line(1,0){1}}  
\put(1,0){\framebox(3,5)[c]{$k_2$}}  
\put(4,1){\line(1,0){2}}  
\put(4,4){\line(1,0){2}}  
\end{picture}}  
\put(6,0){\begin{picture}(11, 5)  
\put(0,0){\thicklines\framebox(9,5)[c]{$e^{c_3\frac{i}{2}\sigma_z^1\sigma_z^2}$}}  
\put(9,1){\line(1,0){2}}  
\put(9,4){\line(1,0){2}}  
\end{picture}}  
\put(17,0){\begin{picture}(5, 5)  
\put(0,0){\framebox(3,5)[c]{$k_y$}}  
\put(3,1){\line(1,0){2}}  
\put(3,4){\line(1,0){2}}  
\end{picture}}  
\put(22,0){\begin{picture}(11, 5)  
\put(0,0){\thicklines\framebox(9,5)[c]{$e^{c_2\frac{i}{2}\sigma_z^1\sigma_z^2}$}}  
\put(9,1){\line(1,0){2}}  
\put(9,4){\line(1,0){2}}  
\end{picture}}  
\put(33,0){\begin{picture}(7, 5)  
\put(0,0){\framebox(5,5)[c]{$k_xk_y^\dag$}}  
\put(5,1){\line(1,0){2}}  
\put(5,4){\line(1,0){2}}  
\end{picture}}  
\put(40,0){\begin{picture}(11, 5)  
\put(0,0){\thicklines\framebox(9,5)[c]{$e^{c_1\frac{i}{2}\sigma_z^1\sigma_z^2}$}}  
\put(9,1){\line(1,0){2}}  
\put(9,4){\line(1,0){2}}  
\end{picture}}  
\put(51,0){\begin{picture}(7, 5)  
\put(0,0){\framebox(5,5)[c]{$k_1k_x^\dag$}}  
\put(5,1){\line(1,0){1}}  
\put(5,4){\line(1,0){1}}  
\end{picture}} 
\end{picture} 
\end{center} 
where each nonlocal block $e^{c_j\frac{i}{2}\sigma_z^1\sigma_z^2}$ 
is simulated as shown in the circuit of
Proposition~\ref{proposition:universal}.
It is clear that overall we only need to apply the gate  
$U_f$ at most $6$ times, in order to simulate an arbitrary two-qubit 
 operation. We thereby obtain an {\em upper bound} for  
the applications of the given entangling gate $U_g$ 
to construct an exact universal quantum 
circuit. 
The value of this upper bound depends only
  on the nonlocal   part of the given gate.  
For example, for a Controlled-PHASE gate $C_\phi$ with parameter $\phi\in
[\frac{\pi}2, \pi]$, it takes 6 applications of $C_\phi$
and 7 local gates to simulate any arbitrary two-qubit operation.
However, when $\phi\in (0, \frac{\pi}2)$, we first need to apply $C_\phi$ $n$
times until $n\phi \ge \frac{\pi}2$. Consequently, in this case it takes $6n$
applications of $C_\phi$ and $6n+1$ local gates to implement any
arbitrary two-qubit operation.
Furthermore, this upper bound is uniform in the sense that no 
  matter which two-qubit unitary operation is to be implemented, we can always 
construct a quantum circuit to simulate this operation with applications of 
  the given entangling gate $U_g$ not exceeding the upper bound.

It is instructive to compare these results with the numerical solution
for construction of CNOT obtained in~\cite{Bremner:02}. For the
gate $U_g=e^{\frac{\pi}3 \frac{i}2 \sigma_z^1\sigma_z^2}$,
both procedures need only two applications to obtain the CNOT gate. 
When $U_g=e^{\frac{\pi}5 \frac{i}2 \sigma_z^1\sigma_z^2}$, 
our uniform construction requires four applications, 
whereas the procedure of~\cite{Bremner:02} only needs three
applications to get CNOT. This difference derives from the fact that
our procedure provides a uniform solution and
is not optimized for any specific gate, whereas the procedure
of~\cite{Bremner:02} is near optimal for CNOT. This comparison reveals 
that the uniform property and optimality cannot necessarily be satisfied
simultaneously. 

Our final analysis concerns the efficiency of these analytic circuits. 
We first show that
our basic building blocks of the quantum circuit, namely 
the gates $e^{\gamma\frac{i}2  
  \sigma_z^1\sigma_z^2}$, are locally equivalent to  
the Controlled-Unitary (Controlled-$U$) gates. 
 
\begin{proposition}  
\label{proposition:CU}  
Consider an arbitrary single-qubit gate   
$U=\exp\{i\gamma\, \hat{n}\cdot \overrightarrow{\sigma}\}$, 
where $\gamma\in \R^+$, $\hat{n}=(n_x, n_y, n_z)$ is a unit vector in $\R^3$,  
  and $\overrightarrow{\sigma}$ denotes the  
vector $(\sigma_x, \sigma_y, \sigma_z)$ of Pauli matrices.  
The corresponding Controlled-$U$ gate can be simulated  
by the following quantum circuit: 
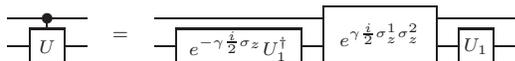
\setlength{\unitlength}{0.15cm}  
\begin{center}  
\scriptsize  
\begin{picture}(45,5)  
\put(0,0){\begin{picture}(7, 5)  
\put(0,1.5){\line(1,0){2}}  
\put(0,4){\line(1,0){7}}  
\put(3.5,4){\line(0,-1){1}}  
\put(3.5,4){\circle*{1}}  
\put(2,0){\framebox(3,3)[c]{$U$}}  
\put(5,1.5){\line(1,0){2}}  
\end{picture}}  
\put(7,0){\begin{picture}(6, 5)  
\put(0,0){\makebox(6,5)[c]{$=$}}  
\end{picture}}  
\put(13,0){\begin{picture}(32,5)  
\put(0,1.5){\line(1,0){2}}  
\put(2,0){\framebox(11,3)[c]{$e^{-\gamma\frac{i}{2}\sigma_z}U_1^\dag$}}  
\put(13,1.5){\line(1,0)2}  
\put(15,0){\framebox(10,5)[c]{$e^{\gamma \frac{i}2\sigma_z^1\sigma_z^2}$}}  
\put(0,4){\line(1,0){15}}  
\put(25,4){\line(1,0){7}}  
\put(25,1.5){\line(1,0){2}}  
\put(27,0){\framebox(3,3)[c]{$U_1$}}  
\put(30,1.5){\line(1,0){2}}  
\end{picture}}  
\end{picture}  
\end{center}  
where  
\begin{equation}  
\label{eq:7}  
U_1=\left\{\begin{array}{cl}  
{\footnotesize \left(\begin{matrix}  
i\sqrt{\frac{1-n_z}{2}}&\sqrt{\frac{1+n_z}{2}}\\  
\frac{n_y-n_xi}{\sqrt{2(1-n_z)}}&  
\frac{n_x+n_yi}{\sqrt{2(1+n_z)}}\end{matrix}\right)},  
&\text{for } n_z\neq \pm 1;  \vspace{0.2cm} \\   
\sigma_x,&\text{for } n_z=1;\\  
I, &\text{for } n_z=-1.  
\end{array}\right. \nonumber 
\end{equation}  
\end{proposition} 
This proposition can be proved by substitution of $U_1$, followed by
direct algebraic computation~\footnote{See EPAPS Document
  No. E-PRLXXX-00-000000 for proof.
A direct link to this document may be found in the online article's HTML
reference section. This document may also be reached via
the EPAPS homepage (http://www.aip.org/pubservs/epaps.html) 
or from ftp.aip.org in the directory /epaps/.
See the EPAPS homepage for more information.}. 

We saw above that all the gates $U_f=e^{\gamma\frac{i}2 \sigma_z^1\sigma_z^2}$  
with  $\gamma\in [\frac\pi{4}, \frac\pi{2}]$ have the 
same upper bounds.
When $\gamma=\frac\pi{2}$,   $U_f$ is locally  
equivalent to the CNOT gate.
From~\cite{Zhang:02}, there exists a one-to-one 
map from the local equivalence classes of  
Controlled-$U$ gates to the points in the interval $[0, \frac{\pi}2]$.
Furthermore, those gates having the same upper bounds as 
the CNOT gate constitute half of this interval, namely $[\frac\pi{4},
\frac\pi{2}]$. Therefore, using the length of the interval as a measure, 
we conclude that {\em exactly half} of the Controlled-$U$ gates can be used to   
construct universal quantum circuits that satisfy the same upper
bound, {\it i.e.}, they can be used just as efficiently as the 
 CNOT gate.
(Note that this is true despite the fact that CNOT is the only 
Controlled-$U$ gate providing perfect entanglement~\cite{Zhang:02}.)
Consequently, there is no need to restrict practical 
 studies of physical implementation of quantum circuits for universal  
computation or for quantum simulations to the standard model 
of CNOT with local gates. 

In summary, we have provided an analytic approach to construct a  
universal quantum circuit that can simulate any arbitrary two-qubit  
operation given any entangling gate $U_g$ supplemented with local gates.
Closed form solutions have been derived for each step   
in this explicit construction procedure. 
The procedure was illustrated on a physical example of simulation of
a solid state spin system with neutral atoms in an optical lattice.  
Our approach provides a uniform upper bound for the applications of
the given entangling gate $U_g$.   
It was found that precisely half of all the
Controlled-$U$ gates have the same uniform upper bounds as the CNOT
gate, {\it i.e.}, they are equally efficient as the CNOT gate for providing  
realizable implementation of arbitrary two-qubit operation.  
This offers new options for realization of interactions in simulation 
of one quantum many-body system by another, as well as for efficient 
implementation of quantum computation.
  
We thank the NSF for financial support under ITR Grant No.  
EIA-0205641 (SS and KBW). KBW thanks the Miller Institute for Basic 
Research for a Miller Research Professorship 2002-2003. 
\bibliographystyle{apsrev}  

\end{document}